\documentclass[mathleft
]{an}
\usepackage{graphicx}
\usepackage{times}
\begin{document}

% The following seven commands are intended for editorial usage and should be ignored by
% the author(s).
\Pagespan{526}{535}% Document's page range. 
% If second parameter is left empty, the last page is computed automatically.
\Yearpublication{2010}%
\Yearsubmission{2010}%
\Month{05}%   
\Volume{331}%  
\Issue{5}% 
\DOI{10.1002/asna.201011365}
\renewcommand{\labelitemii}{$\rightarrow$}

\title{Metallicity and kinematical clues to the formation of the Local Group}

\author{Rosemary F.~G.~Wyse\inst{1}\fnmsep\thanks{Corresponding author:
  \email{wyse@pha.jhu.edu}\newline}
%Example 
%for footnote, note the usage of the \texttt{fnmsep}
%command as separator between institute number and footnote mark} 
%\and  G.H. Ostwriter\inst{2,3}
}
\titlerunning{Formation of the Local Group}
\authorrunning{R.F.G.~Wyse}
\institute{$^1$ Johns Hopkins University, Department of Physics \& Astronomy, Baltimore, MD 21218, USA}

\received{30 May 2005}
\accepted{11 Nov 2005}
\publonline{later}


\abstract{The kinematics and elemental abundances of resolved stars in the nearby Universe can
be used to infer conditions at high redshift, trace how galaxies
evolve and constrain the nature of dark matter. This approach is
complementary to direct study of systems at high redshift, but I will
show that analysis of individual stars allows one to break
degeneracies, such as between star formation rate and stellar Initial
Mass Function, that complicate the analysis of unresolved, distant galaxies. 
}

\maketitle

\section{Introduction}

These are exciting times to study local galaxies, due to a convergence of capabilities:

\begin{itemize}

\item Large observational surveys of individual stars in Local Group
galaxies are feasible, using wide-field imagers and multi-object
spectroscopy, complemented by space-based imaging and spectroscopy, 
with in the near-future the astrometric satellite Gaia providing  6-dimensional phase-space information

\begin{itemize}\item  Imaging surveys need matched spectroscopic surveys for full physics
\end{itemize}
\item High-redshift surveys are now starting to quantify the stellar populations and morphologies of galaxies at high look-back times, equal to the ages of old stars nearby

\item Large, high-resolution simulations of structure formation are allowing predictions of Galaxy formation in a cosmological context

\end{itemize}

I will focus on the first of these capabilities, which provides the
data that enables the fast-growing field of Galactic Archaeology. In
this approach, analysis of the properties of low-mass old stars in our
own Galaxy, and in nearby galaxies, allows us to do cosmology,
locally.  There are copious numbers of stars nearby that have ages
greater than $\sim 10$~Gyr: these formed at look-back times
corresponding to redshifts $ > 2$, and for a subset, perhaps as early as
the epoch of reionization.  The `fossil record' is comprised of the
conserved quantities that reveal conditions at the time the star
formed, such as the chemical abundances in the interstellar medium.
This is a complementary approach to direct study of galaxies at high
redshift; the two approaches to understanding galaxy evolution analyse
on the one hand snapshots of different galaxies at different times
(redshifts) and on the other hand the evolution of one galaxy.  A
significant advantage of Galactic Archaeology, studying resolved
stars, is that one can derive {\it separately} the stellar metallicity
distributions and age distributions, and constrain variations in the
massive-star Initial Mass Function (IMF) from elemental ratios. One
can then break degeneracies that hamper the interpretation of
photometry and even spectra of the integrated light of galaxies,
including the well-known age--metallicity case and
star-formation-rate--IMF.  Analysis of the motions of stars within a given galaxy can also provide the mass profile of the underlying potential, going from kinematics to dynamics.

The key questions that can be addressed, and which I will touch on in this review, include

\begin{itemize}
\item How do galaxies form? 
\begin{itemize}
\item star formation histories
\item   mass assembly histories
\item  link to black hole growth
\item  the physics of `feedback'
\end{itemize}
\item  What is the nature of dark matter?
\begin{itemize}
\item   determines potential well shape
\item  determines merger histories
\end{itemize}
\end{itemize}

\section{Testing the $\Lambda$CDM paradigm of structure  formation}

The standard model of structure formation is that of $\Lambda$CDM,
where most of the gravitating material in the Universe is in the form
of cold dark matter, and most of the energy is in a dark component
that behaves as a constant density Cosmological Constant. This has
proven to be an excellent description of the Universe on large scales,
probed by the Cosmic Microwave background (e.g.~Komatsu et al.~2009),
and the large-scale power-spectrum of the distribution of galaxies
(e.g.~Percival et al.~2010).  Gravity dominates on these scales, so that hot and cold dark matter fit equally well, and
the physics of the constituents of dark matter is expected to be
manifest only on smaller scales, such as sparse galaxy groups and individual galaxies 
(e.g.~Ostriker \& Steinhardt 2003).  These are indeed the scales on
which cracks are appearing in the $\Lambda$CDM armour (e.g.~Wyse 2001; Peebles \&
Nusser 2010).

Structure formation in the $\Lambda$CDM paradigm is hierarchical, with
smaller scales forming first, due to the shape of the power spectrum
of primordial density fluctations. This results in galaxy formation
and evolution being largely driven by mergers. The outcome of a merger
between two galaxies~(i.e. dark matter haloes with associated baryonic
material bound within them) depends in large part to the relative
fractions of dissipationless material (dark matter, stars) and dissipational
material (gas -- ignoring dissipational dark matter models) 
(e.g.~Tinsley \& Larson 1979; Zurek, Quinn \& Salmon 1988; Governato
et al.~2009), the assumed physical conditions and equation of state of
the dissipational material.  The mass ratio  and relative densities of the merging systems are also extremely important. 
Simulations that are dark-matter-only have the simplest physics, and
-- following on from the pioneering work of Moore et al.~(1999) and of Klypin et al.~(1999) -- the highest resolution realizations for a Milky Way galaxy analogue
show significant merging at all redshifts, and persistent substructure on
all mass scales throughout the final, zero redshift, galaxy
(e.g.~Stadel et al. 2009). Substructure makes its presence felt
through its gravitational interactions and assimilation through
mergers, which can lead to heating of initially cold stellar
components (e.g. Quinn \& Goodman 1986; Kazantzidis et al. 2009). 
Thin disks are very susceptible to the destructive effects of mergers (of roughly equal mass),
and in the $\Lambda$CDM scenario the disks we see today largely formed after merging ceases to be
active, which is typically a redshift $ z \sim 1$.  The possibility 
that gas accretion into dark halos is dominated by cold streams, rather than the steady accretion of gas initially shock-heated to the halo virial temperature (White \& Rees 1978), leads
to the creation of fat, clumpy, turbulent disks of gas and stars at
higher redshifts, $z \sim 2$ with active star formation (Ceverino,
Dekel \& Bournaud 2010), evolving into quiescent spheroid-dominated
systems at lower redshift.  The initial formation at $z \sim 2$ of thin disks -- that are later destroyed through merging -- is seen in high-resolution 
multi-phase hydrodynamic simulations within the $\Lambda$CDM framework
(e.g.~Scannapieco et al.~2009). The relationship of these predicted early thick and thin disks to the star-forming galaxies observed at redshifts $\sim 2$ (e.g.~Genzel 2009) is as yet unclear, but intriguing. 

A thin stellar disk is likely to be significantly affected
by even very unequal mass-ratio mergers (minor mergers), with a significant fraction of the
pre-merger orbital energy being absorbed into the internal degrees of
freedom of the disk, resulting in heating, and creation of a thick
stellar disk.  Gas of course can radiate energy and cool, but the
stellar component of the disk cannot.  Adiabatic compression of the
heated stellar disk,by gas settling later into the midplane,results in
both a decrease of scale height of the stellar component and an increase of its velocity dispersion
(e.g.~Toth \& Ostriker 1990; Elmegreen \& Elmegreen 2006).  Some of the
orbital angular momentum will also be absorbed, in general resulting in a tilt of
the disk plane.  The relative density profiles and the details of the
satellite's orbital parameters also are important determinants of the final
outcome, in terms of the stellar disk structure (and of the fate of the satellite).

The robustness, masses and orbital parameters of substructure in
$\Lambda$CDM, and the typical merging history of large galaxies, leads
to the expectation that thick disks should be extremely common, and
that the heating of thin disks should continue to late times. A vivid
illustration of the results, 
using a pure N-body code and a cosmologically motivated retinue of
satellite galaxies impinging on a stellar thin disk, assumed in place
at redshift $z = 1$, are shown in Kazantzidis et al.~(2009).  As noted in several papers, in the concordance
$\Lambda$CDM cosmology, a typical dark halo of the mass of that
inferred for the Milky Way ($\sim 10^{12}$M$_\odot$) will have suffered significant merging with
relatively massive satellites (up to $10^{11}$M$_\odot$) since a
redshift of unity (e.g.~Stewart et al.~2008; Fakhouri, Ma \& Boylan-Kolchin, 2010).  While the presence of
gas may prevent the {\it destruction\/} of an existing thin disk, the
{\it heating\/} is unavoidable, leaving an imprint in the stellar velocity dispersions
(which as noted above, only increase with subsequent adiabatic growth of an
embedded gas-rich thin disk).

Material is stripped from satellites as they merge, depending to first
order on the relative density compared to that of the larger galaxy
interior to the satellite orbit.  The less dense, outer parts of
satellites are thus stripped more easily, at larger distances from the
central regions of the large galaxy.  The typical orbits of merging
substructure are elliptical, albeit with the eccentricity distribution
peaking at close-to-parabolic orbits (e.g.~Benson 2005; Khochfar \&
Burkert 2006).  Debris from a given satellite that survives to the
disk plane is broadly expected to be distributed in a thick torus, with radial extent indicating the location at which the satellite was disrupted, and 
some          stars in the debris could have orbital parameters similar to the
heated thin disk, now the thick disk (e.g.~Huang \& Carlberg 1997),
and even to the surviving old thin disk (Abadi et al.~2003).  The dark matter
stripped from satellites during merging also builds up a dark-matter
disk (Read et al.~09), the mass of which is severely constrained, at
least locally, by the vertical motions of stars, which are consistent
with no significant disk dark matter (the `$K_Z$' analysis and Oort
limit, Kuijken \& Gilmore 1989, 1991; Holmberg \& Flynn 2004;
Bienaym{\'e} et al.~2006). Torques during a minor merger will drive
gas and stars inwards, to contribute to the inner disk and bulge
(e.g.~Mihos \& Hernquist 1996).

In $\Lambda$CDM, the stellar halo of disk galaxies is postulated to be
formed from disrupted satellite galaxies, with structure in coordinate
space persisting for many dynamical times, even to the present times
in the outer regions, where timescales are longest. Graphic
illustration of the predicted structure, should the stellar halo consist entirely of accreted satellites,  is given in Johnston et
al.~(2008; note that this is not a fully self-consistent model). 
Some fraction of the stars now in the inner halo/bulge
may be disrupted disks from earlier stages of sub-halo merging, giving
a `dual halo' (e.g.~Zolotov et al.~09).

\subsection {The Milky Way and M31 as templates}

As noted above, the stellar halo, bulge, thick disk and even some part
of (old?) thin disk of a typical large disk galaxy is predicted, in $\Lambda$CDM,  to be
created through the effects of mergers. One should see signatures of these origins in
the stellar populations, since some memory of the initial conditions
when a star is formed is retained.  The Local Group provides an ideal
test-bed of such predictions.  Observational effects include
\begin{itemize}
\item Coordinate space structure
\item  Kinematic (sub)structure
\item  Chemical abundance signatures
\item  Distinct age distributions
\item  Properties of surviving satellite galaxies 
\end{itemize}

Kinematic phase-space structure should survive longer than
will coordinate-space structure, and chemical signatures of distinct
populations are the most long-lived.  Kinematics and chemical distributions are most robustly
determined  through spectroscopy.

Stars today probe conditions at a look-back time equal to their age;
10Gyr corresponds to a redshift $z \sim 2$ for the concordance
$\Lambda$CDM cosmology. Such stars, represented by main sequence, red
giant stars and horizontal branch stars, are accessible in the Local
Group with current capabilities.  Only young, massive stars may be
studied currently in galaxies beyond the Local Group; the wealth of
information that can be derived from these was described by Rolf Kudritzki in his
Schwarzschild lecture; an Extremely Large telescope is required to
reach lower mass stars beyond the Local Group.

The Local Group consists of two large disk galaxies, M31 and the Milky
Way galaxy, plus a low luminosity, very late-type disk galaxy, M33, a
companion to M31. There are also retinues of gas-rich dwarf irregular (dIrr) 
and gas-poor dwarf spheroidal (dSph) satellite galaxies of each of the large
disk galaxies, with many recent  discoveries (discussed further
below).  We will discuss how photometric and (in particular)  
spectroscopic surveys of these galaxies reveal the important processes
by which galaxies evolve.

\section{Photometric surveys}

\subsection{The Milky Way galaxy}

The first irrefutable evidence for ongoing merging between the Milky
Way and its satellite galaxies was provided by the (serendipitous)
discovery of the Sagittarius (Sgr) dSph galaxy, by Ibata, Gilmore \& Irwin (1994). The coordinate-space overdensity of the core of this galaxy is illustrated in Fig.~1 of Wyse, Gilmore \& Franx (1997). 
This galaxy was detected kinematically during  a
spectroscopic study of the bulge of the Milky Way (Ibata \& Gilmore
1995). Member stars of this `moving group' also have a redder colour
distribution than the field bulge stars, a manifestation of their
distinct age and metallicity distributions (determined by subsequent
observations e.g.~Layden \& Sarajedini 2000; McWilliam \& Smecker-Hane
2005; Sbordone et al.~2007; Siegel et al. 2007; Chou et al.~2010). The
Sgr dSph contains no detectable gas, down to
limits of $\sim 140$~M$_\odot$ of atomic Hydrogen (Grcevich \& Putman
2009 and refs.~therein), and is clearly dark-matter dominated, based on the velocity dispersion of its member stars 
(e.g.~Ibata et al.~1997).  This lack of gas and dominance by dark matter are characteristic of the dSph satellite galaxies.  

\begin{figure}
\includegraphics[width=83mm]{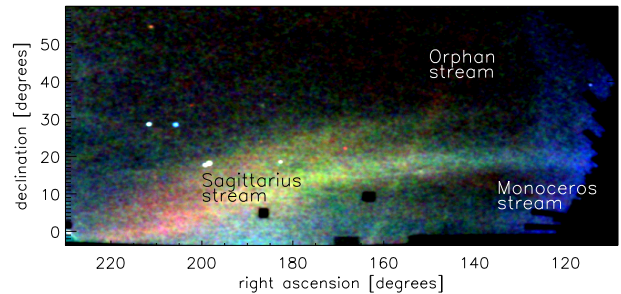}
\caption{The `Field of Streams' -- and dots. This annotated version is
courtesy of Gerry Gilmore and Vasily Belokurov (see Belokurov et al.~2006). The figure shows the
distribution of main sequence turnoff stars on the sky, selected from
the SDSS imaging data by the colour and magnitude cuts $19.0 < r <
22.0$ and $g -r < 0.4$, colour-coded in the figure by apparent
magnitude, i.e.~by distance from the Sun. Blue represents the nearest
stars (at $\sim 10$~kpc), then green, then red (at $\sim 30$~kpc). The distribution of stars is clearly not uniform or smooth.}
\label{FOS}
\end{figure}

A new vitality was injected into the field by the advent of the Sloan
Digital Sky Survey (SDSS) imaging data. The uniformly excellent
photometry across a large fraction of the Northern sky revealed a very non-uniform distribution of resolved stars, dubbed the  `Field of
Streams' (Belokurov et al.~2006; see Fig.~\ref{FOS} here). 

The distribution of stars in the Northern sky shown in Fig.~\ref{FOS} is
dominated by the two swathes of stars in the tidal streams from the
Sgr dSph. Tidal debris from this system appears to be manifest not
only in these two wraps (e.g.~Fellhauer et al.~2006) but also as a significant part of the
Virgo overdensity (e.g.~Prior, Da Costa \& Keller 2009). The outer stellar halo, at Galactocentric distances of greater than
$\sim 15$~kpc and traced out to $\sim 60$~kpc,  shows significant coordinate space structure (Bell et
al.~2008; Watkins et al.~2009), the vast majority of which is associated with the Sgr
dSph, and to a lesser extent with the Hercules-Aquila Cloud (Belokurov et al.~2007a), accounting for perhaps 10\% of the total mass of the stellar
halo (M$_{star,halo} \sim $ $10^9$~M$_\odot$).

\begin{figure}
\includegraphics[width=60mm,angle=270]{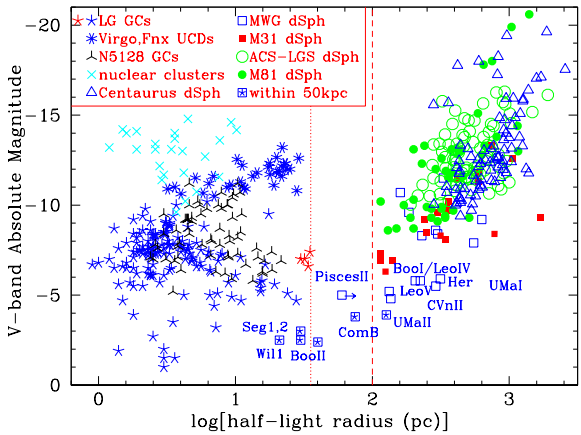}
\caption{Distribution of star clusters and dwarf galaxies in the plane
of stellar half-light radius {\it vs\/} absolute V-band magnitude. The
faint systems discovered in the SDSS imaging data that have been
reported to be galaxies are identified by name.  There are only three
systems between the vertical red lines (dotted on the left, at a
half-light radius of 36~pc, and dashed on the right, at a half-light
radius of 100~pc).}
\label{Mv_rh}
\end{figure}

 The numerous `dots' are also of extreme interest: these are 
gravitationally bound satellites of the Milky Way, ranging in
luminosity from only a few hundred solar luminosities to millions of
solar luminosities.  Whether they are star clusters or satellite
galaxies is only revealed by spectroscopic measurements of their
internal kinematics, from radial velocities of member stars --
galaxies, by definition, are held together by the gravity of 
dark matter haloes, while star clusters are baryon-dominated
self-gravitating systems; this is discussed further below.  Photometry
alone provides distances -- from the colour-magnitude diagram (CMD) of member stars --  the
scale-lengths of the stellar distribution, and the total
luminosity. These result in the size--absolute magnitude plot shown in
Fig.~\ref{Mv_rh}, which includes known star clusters and dwarf
galaxies (and is an annotated update of Fig.~1 in Gilmore et
al.~2007).

The unoccupied region at the lower right is a selection effect -- too
few stars over too large an area -- and the boundary approximately follows a line of
constant V-band surface brightness equal to $\sim 31$~mag/sq~arcsec
(see Belokurov et al.~2007b).  There is a noticeable gap, emphasised in
the Figure by the vertical red lines. As the SDSS imaging data have
allowed lower and lower luminous systems to be discovered, they have
largely been on either side of the gap (the newly discovered faint
globular clusters are unlabelled).  Indeed only three systems fall in
the gap.  Of these, two are closer than 50~kpc, within the realm of
expected tidal effects -- manifest in the Sgr streams -- and one,
Pisces~II, is surrounded by a `sea' of blue horizontal branch stars
that were not included in the estimation of the half-light radius
(Belokurov et al.~2010), and so will likely move out of the gap, to
larger sizes, with deeper photometry.  The nearby systems, within
50~kpc, are indicated in Figure~\ref{Mv_rh} and these tend to be
rather elliptical and/or distorted (e.g.~Martin, de Jong \& Rix 2008), which may well
indicate the effects of tides.  Their present structure cannot be
assumed to be the equilibrium structure at formation.

\subsection{M31}

This subsection should perhaps be labelled `M31 / M33', since  deep
imaging surveys are revealing hints of low surface brightness stellar features
between the two,  which, together with the wide-spread HI gas (Putman et al.~2009) are suggestive of past interactions.  The most comprehensive
wide-field survey to date is that of the PAndAS collaboration (McConnachie et al.~2009),
building upon the pioneering work of Ibata et al.~(2001) and Ferguson et al.~(2002).  Evolved
stars in either of the hydrogen-shell burning phase (red giants) or core-helium
burning phase (red clump and horizontal branch) are fairly
straightforward to image at the distance of M31, at least in the lower
surface-brightness  outer disk
and halo (avoiding source crowding and confusion).  Similarly to the dominant stream from the Sgr dSph in the
Milky Way, the structure in the stellar halo of M31 is dominated by the `Giant Stream' (Ibata et al.~2004), together with significant other structure (e.g.~Tanaka et al. 2010). Not only
the Giant Stream but also smaller-scale structure can
result from one accreted system (e.g.~Fardal et al.~2008), but as in the Milky Way the situation is complex and many merged satellites may be involved (Font et al.~2008).

Again, many new satellites of M31 have been discovered by deep imaging
surveys (e.g.~Martin et al.~2009 and references therein); these are
included in the satellite systems shown in Fig.~\ref{Mv_rh}; the M31 satellites also
avoid the size gap.   One
might note that the earlier finding of significantly larger radii of M31
satellites compared to those of the Milky Way (McConnachie \& Irwn
2006) is not held up by the fainter systems.

The colour of a red giant star of a given luminosity is more sensitive
to chemical abundance than to age.  Counts of red
giants divided into two categories, red and blue, reveal significant
chemical inhomogeneities across the outer disk and halo of M31
(Ferguson et al.~2002).  The mean metallicity of the dominant `halo'
(really `non-thin-disk') component of M31, derived from colours of the
RGB stars, is significantly higher than that of the stellar halo of
the Milky Way, being [Fe/H] $\sim -0.6$~dex (cf.~Mould \& Kristian
1986; Durrell, Harris \& Pritchet~2004).  This is equal to the mean metallicity of the
Galactic thick disk at the solar neighbourhood, leading to speculation
that these stars represent the thick disk of M31, and that the Galactic thick disk, rather than the halo,  should be called `Population II' to match the population in M31 that Baade resolved into stars (Wyse \& Gilmore 1988). 
This high a value of mean chemical abundance, combined with the inferred old age from the CMD, implies
that the stars formed within a fairly deep potential well,
significantly more massive than a typical dwarf galaxy.

A large, high S/N spectroscopic survey is required to provide the kinematics and robust metallicities of both the field stars and satellite galaxies. This will allow testing of models for interactions between M31 and its satellites, including M33, and masses estimates for the surviving satellites.  These have been initiated with 8m-class telescopes (e.g.~Chapman et al.~2008; Letarte et al.~2009; Gilbert et al.~2009; Kalirai et al.~2010).

\section{Spectroscopic surveys}

While there exist photometric techniques, such as the colour of the
RGB with an assumed age, or the UV excess of main sequence F/G stars
(measured either by broad-band filters or Str{\"o}mgren photometry),
to determine chemical abundances, and other photometric techniques for  effective
temperatures, spectroscopy remains the most robust and reliable means
to determine stellar parameters, in particular the elemental
abundances.  Spectroscopy is also critical for determining kinematics
for stars, particularly those far enough away that proper motions are
too small to be measured (in the era prior to Gaia!)

There is a need for a variety of spectroscopic surveys at different
spectral resolutions. Moderate resolution, here defined as a resolving
power of a few thousand at optical wavelengths, provides spectra that
are sufficient to determine line-of-sight motions to better than
10~km/s and metallicities to $\sim 0.2$~dex. With statistically large
samples of stars these data can then be used to define the
distribution functions of the major stellar populations under study,
and investigate decompositions into different components. An
individual star can then, in a probabilistic manner, be assigned to a
given component.  A significant amount of physics lies in the detailed
shape and tails of the distribution functions, and these require
large samples with well-defined selection functions.

High resolution, here defined as a resolving power of several tens of
thousand at optical wavelengths, provides spectra that are sufficient
to determine line-of-sight motions to much better than 1~km/s and
individual elemental abundances to $\sim 0.05$~dex.  This precision in
the velocities allows the mapping of cold stellar substructure, such
as tidal streams, and the internal kinematics of low luminosity
satellite galaxies.  Ideally one would have both high and moderate
resolution spectra for stars drawn from the same parent sample, such
as is possible with the RAVE spectroscopic survey (e.g.~Steinmetz et
al.~2006), as discussed below.

\subsection{Masses and mass profiles of dwarf galaxies}

Precise measurement, with echelle spectroscopy, of the line-of-sight
velocities of member stars of dSph satellite galaxies was
pioneered by Marc Aaronson, who derived a high mass-to-light ratio for
the Draco dSph, based on velocities of three carbon stars.  He
obtained a {\it lower limit\/} to the central velocity dispersion of
6.5~km/s (Aaronson 1983), perfectly consistent with the modern value
of 9~km/s (Walker et al.~2009).  Mass-to-light ratios have been
estimated for the systems in Fig.~\ref{Mv_rh} leading to the
conclusion that all systems to the right of the gap are embedded in
dark matter haloes, i.e.~are galaxies, while all equilibrium systems
to the left of the gap are star clusters.  This then provides the
physical interpretation of the size distribution in terms of a minimum
scale for systems dominated by dark matter (Gilmore et al.~2007).  The
scale in Fig.~\ref{Mv_rh} is that of the light; given that baryons
dissipate binding energy while gaseous, and become self-gravitating to
form stars, this stellar half-light radius is expected to be a lower
limit to the dark-matter scale-length.  Such a minimum scale to dark matter of greater than $\sim 100$~pc is not expected if the dark matter is cold, and may be indicative of warm dark matter, such as sterile neutrinos (e.g.~Kusenko 2009). 

\begin{figure}
\centering
\includegraphics[width=3.22in]{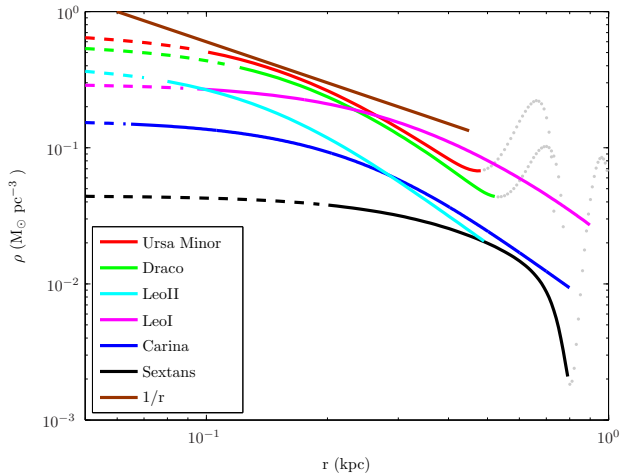}
  \caption{Derived mass density profiles from (isotropic) Jeans-equation
analyses of the stellar velocity dispersion profiles of 6 dwarf
spheroidal galaxies.  Also shown is a $r^{-1}$ profile, to which
$CDM$-mass profiles are predicted to asymptote. The analysis is in each
case reliable out to $r \sim 500$~pc. This figure is taken from
Gilmore et al.~(2007)}
\label{profiles}
    \end{figure}

Derivation of mass profiles, as opposed to estimates of total mass,
requires  velocity information as a function of projected radius,
not just for the central regions.  The most straightforward analysis uses the
Jeans equations, which relate the mass profile to the stellar light
profile through the radial dependences of (line-of-sight) velocity
dispersion and velocity dispersion anisotropy. However, this introduces a
well-known mass--anisotropy degeneracy (Binney \& Mamon 1982),
which can be broken by use of the detailed line-of-sight velocity distribution,  rather than its moments,  such as the velocity 
dispersion (Gerhard 1993; Gerhard et al.~1998).  This latter approach
requires sample sizes of thousands of stars, well-distributed across
the face of the galaxy, but with the central regions remaining
critical, since it is here that the physics of the dark matter particle will be manifest most strongly.  Such datasets are now being acquired (Gilmore et al.~in preparation).  The
spherical Jeans' equation approach, with assumed isotropy, has the
advantage of simplicity and when applied to a set of galaxies can
identify any differences among them. The results of such an
investigation, applied to six of the classical dSph satellites of the
Milky Way is shown in Fig.~\ref{profiles}, taken from Gilmore et
al.~(2007) -- the derived mass density profile for each galaxy favours an
inner core, rather than the cusped profile predicted for cold dark
matter (Navarro, Frenk \& White 1996). Further, all the profiles are very similar.  This similarity
is despite the fact that this set of galaxies covers a wide range in
star-formation histories, implying that an appeal to feedback from
star-formation to smooth out an inner cusp into a core would need to
be tuned to each galaxy. 

\subsection{Elemental abundances: beyond metallicity}

Different elements are created in stars of different masses and
different evolutionary stages, on different timescales. Thus the
pattern of elemental abundances seen in a stellar population reflects
the star formation history and the stellar initial mass function (and
the binary population) that provided the enrichment of the stars we
observe now.  A consequence is that old low-mass stars nearby provide
insight into the stellar IMF at high redshift (look-back times at
least equal to the age of the star).

Core-collapse supernovae, the death throes of
massive, short-lived,  stars, are the primary sources of the $\alpha$-elements,
created in steady-state, pre-supernova nucleosynthesis and ejected in the explosion.  There is little iron produced, since the iron core is photo-disintegrated and largely forms the stellar remnant. 
Type~Ia supernovae are produced by explosive nucleosynthesis in a white dwarf of mass greater than the Chandrasekhar limit, created by accretion from a binary companion. Such systems evolve on timescales that can be
as long as a Hubble time and provide mostly iron, about ten times as much per supernova as a Type II supernova.

Stars formed in the earliest stages of a
self-enriching system will have high levels of [$\alpha$/Fe],
reflecting the products of Type~II (core-collapse) supernovae. There
is negligible range in lifetime for Type~II progenitors, compared to
timescales of interest for star formation,  and if there is good mixing
of the supernova ejecta one expects to see an IMF-averaged yield
of [$\alpha$/Fe] in the next generation of stars.  The most massive
Type~II progenitors produce the highest values of [$\alpha$/Fe]
(Gibson 1998; Kobayashi et al.~2007), so that a massive-star IMF
biased towards the most massive stars will leave a signature, namely even
higher values of IMF-averaged [$\alpha$/fe].  Some (model-dependent)
time after the onset of star formation, white dwarfs will accrete
sufficient material to explode as Type~Ia supernovae. The first Type
Ia are expected to result from progenitors with main sequence mass
just below the threshold for core-collapse ($\sim 8$~M$_\odot$), and
these could in principle evolve to explosion in less than $10^8$~yr in
the double-degenerate model (e.g.~Matteucci et al.~2009). It takes longer for
sufficient Type Ia to explode and for their ejecta to be incorporated
in the next generation of stars, and a typical lag seen in chemical
evolution models is $\sim 1$~Gyr (Matteucci et al.~2009).  Of
course, the {\it iron abundance\/} corresponding to this time depends
on the star formation rate and gas flows in or out.  Inefficient
enrichment, due, for example, to a slow rate of star formation or to
loss of metals by winds, results in the signature of `extra' iron from
Type Ia being seen at lower values of [Fe/H].
High rates of star formation and efficient enrichment will
produce a `Type~II plateau' that extends to higher [Fe/H] (e.g.~Fig.~1 of Wyse \& Gilmore 1993).  One then
expects that systems with low rates of star formation, over extended
periods -- such as the dwarf spheroidal satellites of the Milky Way --  should show low values of [$\alpha$/Fe] at low values of
[Fe/H]  (Unavane, Wyse \& Gilmore 1996). 

\begin{figure}
\includegraphics[width=60mm,angle=270]{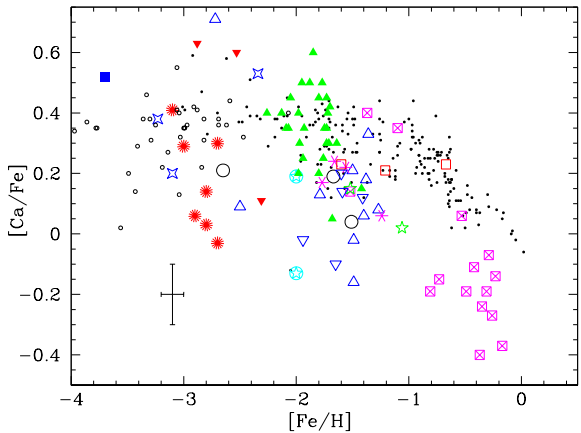}
\caption{Abundance of calcium, an $\alpha$-element, relative to iron,
for field stars in the Milky Way Galaxy (small open and filled black
circles, from Cayrel et al.~2004 and Fulbright 2000, respectively) and
stars that are members of representative dwarf spheroidal galaxies,
satellites of the Milky Way. Typical errors as indicated in the bottom right.  The large black open circles represent
stars in the Ursa Minor dSph (Sadakane et al.~2004), the magenta
squares, enclosing crosses, represent stars in the Sgr dSph (Monaco et
al.~2005), the red many-armed asterisks represent stars in the Sextans dSph (Aoki
et al.~2009), the blue open triangles represent stars in the Carina
dSph (usual orientation, Koch et al.~2008a; upside down, Shetrone et al.~2003), the green filled triangles represent stars in
the Leo II dSph (Shetrone et al.~2009), the open red squares represent stars in the Fornax dSph (Shetrone et al.~2003), the six-armed magenta asterisks represent stars in the Sculptor dSph (Shetrone et al.~2003), the two green open star symbols represent stars in the Leo~I dSph (Shetrone et al.~2003), the two cyan circles,
enclosing star symbols, represent two stars in the Hercules dSph (Koch
et al.~2008b), the blue 4-pointed star symbols represent stars in the
Ursa Major I dSph (Frebel et al.~2009), the red upside-down filled
triangles represent stars in the Coma Berenices dSph (Frebel et al.~2009) and the filled blue square represents a star in the Bo{\"{o}}tes~I dSph (Norris et al.~2010).  These last four dSph were all discovered within the SDSS imaging dataset and are very faint, fainter than absolute magnitude $M_V \sim -7$, well into the regime of globular star clusters (Belokurov et al.~2006, 2007b; Zucker et al.~2006).} 
\label{labelca}
\end{figure}

Observations have confirmed this
expectation.  A compilation of [Ca/Fe] for stars in the Milky Way and in representative gas-poor satellite galaxies is shown in Fig.\ref{labelca}; the stars in satellite galaxies lie systematically below the 
field halo stars at the same value of [Fe/H], leading to the
conclusion that accretion of stars from dwarf galaxies like those
observed today could not play an important role in the formation of
the stellar halo (Venn et al.~2004). Complementary, independent age
information that the vast majority of halo stars are {\it old}, contrasting with the
dominant intermediate age population of the luminous dSph, further
constrains progenitors to consist only of stars formed early, in a short-lived burst of star formation that happened a long time ago (e.g.~Unavane et al.~1996).  A progenitor that would have been expected to have an extended star formation history, similar to a typical dSph, would have had to be accreted $\sim 10$~Gyr ago for its member stars to contribute to the bulk of the stellar halo (Unavane et al.~1996).

A new result is seen too, as observations have pushed to lower and
lower metallicities within each system: we find, within the errors and
limitations of small numbers, a consistent value for the enhanced
`Type II plateau' in [$\alpha$/Fe] in all galaxies for their lowest
metallicity member stars.  These are the stars which one might expect
to have formed in the earliest stages of star formation within each
individual galaxy, and thus to show (pre)enrichment by core-collapse
supernovae only. The fact that there is only moderate scatter in [$\alpha$/Fe] in these stars,
following the field halo stars, implies that (i) the massive-star IMF
is invariant and (ii) well-sampled and the ejecta well-mixed.  These
place rather stringent requirements on the early star formation.  A
further conclusion is that the stellar halo could form from any
system(s) in which star-formation is short-lived, so that only Type II
supernovae have time to enrich the star-forming gas, and enrichment is
inefficient so that the mean metallicity is kept low. These could be
star clusters, galaxies, or transient structures.

The old age and `Type II' elemental abundances of the bulk of the
field halo stars limits any late merging to be of systems that formed
stars early, with short duration; we know of only one such luminous
dwarf galaxy with no detectable age spread,  that in Ursa Minor. Even for this system at least one star shows lower ratios of calcium to iron than does the bulk of the stellar halo, as shown in Fig.~\ref{labelca} (large black open circles; Sadakane et al.~2004). 
Recent models within the framework of $\Lambda$CDM argue that the field halo of the Milky Way formed from 
a few LMC-mass
systems that were accreted early, and, unlike the LMC,  had star-formation truncated
abruptly at that time (e.g.~Robertson et al.~2005).  This has not yet been demonstrated within a self-consistent model of star formation and chemical enrichment. 

\begin{figure}
\includegraphics[width=83mm]{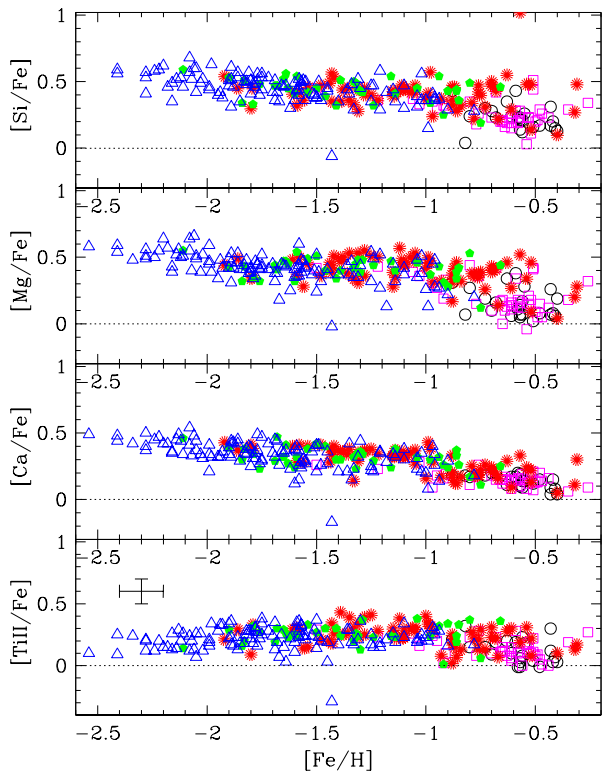}
\caption{Elemental abundances, from analyses of high-resolution
spectra, for stars selected from the RAVE survey database to be low
metallicity but with disk-like kinematics. The distances and
kinematics have been re-calculated based on the stellar parameters
derived from the echelle spectra and isochrone fits.  The blue
triangles represent stars that have high probability to be halo stars,
the red asterisks high probability thick-disk stars, with the green
pentagons having similar probability of belonging to either the halo or thick disk. The black circles represent high probability thin-disk stars and the  magenta open squares are either thin- or thick-disk stars. }
\label{greg_RAVE}
\end{figure}

The stars in the Milky Way in Fig.~\ref{labelca} are not identified by kinematics
or spatial location into distinct stellar components.  When this is
done, further insight can be achieved. Fig.~\ref{greg_RAVE} shows the results from
a large (by the standards of high-resolution spectroscopy) survey of
metal-poor stars, with an emphasis on candidate disk members (Ruchti
et al.~2010).  The sample was selected from the Radial Velocity (RAVE)  spectroscopic
survey (Steinmetz et al.~2006) as having space motions more consistent
with disk than halo, but being of low metallicity; the stellar
parameters from the RAVE pipeline were used in this sample selection.
The RAVE survey is obtaining
moderate-resolution (${\cal R} \sim 8000$) spectra, around the IR
Calcium triplet, for a magnitude-limited sample of bright stars, $I <
12$, using the multi-object fibre spectrograph (the 6dF instrument) on
the UK Schmidt telescope.  The aim is for a final sample of 1~million
stars, allowing for a statistically significant sampling of the
kinematic and metallicity distributions of each of the stellar
components of the Milky Way.  These stars are bright enough that
follow-up echelle spectra of `interesting' subsamples (such as metal-poor disk
stars) can be fairly easily obtained on 4m- or 8m-class telescopes,
while still probing distances of several kiloparsec from the solar
neighbourhood (using giant stars).

The stars with elemental abundances shown in Fig.~\ref{greg_RAVE} have been
re-classified into halo, thick disk and thin disk on the basis of the
stellar parameters derived from the echelle observations, together
with isochrone fitting for improved estimates of the gravity 
(particularly important for the distance estimates for low gravity,
luminous red giant stars; we estimate that our distances for giants are accurate to 20-30\%).  The stars in the RAVE catalogue are sufficiently bright
that proper-motion measurements are available, allowing full 
space motions to be derived once distance are estimated (the RAVE
radial velocities are accurate to better than a few km/s).  The
component assignment is probabilistic by necessity, since the
components have overlapping properties -- and indeed characterising,
then understanding, that overlap is a major science goal of the full RAVE survey.  The different
population assignments are indicated by different symbols in the
Figure (see Ruchti et al.~2010 for details of the range of criteria
used in the assignments). The elemental abundances were derived
following the formalism of Fulbright (2000).

It is evident that the thick disk extends at least to [Fe/H] $\sim
-1.75$~dex, and that these stars show the same enhanced values of
[$\alpha$/Fe] as do halo stars, for a range of $\alpha$-elements.
Most of the stars classified as thick disk or halo are giants, and probe distances of several kiloparsec from the Sun, the first time that elemental abundances for the metal-poor thick disk has been measured at these distances. 
Similar elemental abundance results for smaller samples of (high-velocity) thick disk and halo (dwarf) stars at
[Fe/H] $\sim -1$~dex were reported previously (Nissen \& Schuster
1997), and also for a local sample (distances less than 150~pc) of dwarf stars for the full range in 
[Fe/H] of (Reddy \&  Lambert 2008; Reddy, Lambert \& Allende-Prieto 2006).  The enhanced values of [$\alpha$/Fe]
imply that core-collapse supernovae dominated the enrichment of the
gas clouds from which these stars formed, which in turn implies that
the stars formed within a short time after the onset of star
formation. It could still be possible to incorporate an age range
greater than $~\sim 1$~Gyr, should this be established by other means,
if the stars were to form in distinct sub-units which each had a short
duration of star formation, but different times of onset, but this
requires some fine-tuning (see Gilmore \& Wyse 1998).  A simpler
interpretation is that there is a narrow age range and that the
metal-poor thick disk (and halo) stars formed during a short-lived
epoch of star formation.

The similarity of the elemental abundances across the halo/thick disk
transition implies that the core-collapse supernovae that pre-enriched
the stars came from a very similar massive-star Initial Mass Function.  Further, the low scatter implies good sampling of the massive-star IMF and good mixing within the interstellar medium. This would seem to imply large star-forming regions. 

Several models for the thick disk invoke a minor merger and associated
heating of a pre-existing thin disk; one then expects to find stars
above the disk plane that came from the previous thin disk, together
with stars from the now-shredded satellite that merged (e.g.~Gilmore,
Wyse \& Norris 2002).  Distinguishing these two contributions relies
upon stellar population signatures.  It is clear that the bulk of the
thick-disk stars near the solar neighbourhood, for which the mean
metallicity is relatively high, [Fe/H] $ \sim -0.6$~dex, and the mean
age old, $\sim 10-12$~Gyr, must have formed within a fairly deep
potential well, since even the Large Magellanic Clouds did not
self-enrich to this level until a few Gyr ago.  The metal-poor end of
the thick disk could more plausibly be associated with debris from
accreted dwarf galaxies.  The elemental abundance pattern seen in Fig.~\ref{greg_RAVE} 
requires that star formation in any such progenitor be very
short-lived, which, compared to the typical extended star formation
history, and early onset, in surviving dwarf galaxies implies that any such
dwarf galaxy be accreted a long time ago.  This is consistent with
inferences from the old age of stars in the bulk of the thick disk, but not easy to reconcile with the late merging and accretion of satellites predicted in  
$\Lambda$CDM (e.g.~Abadi et al.~2003). 

Radial migration of stars and gas in disks, due to resonance with
transient spiral arms (Sellwood \& Binney 2002), perhaps in concert
with the bar (Minchev \& Famaey 2009) is potentially very
important. Extremely efficient migration, with a probability that is
independent of the amplitude of random motions of the stars, can
create a metal-rich thick disk locally out of the inner regions of the
disk (Sch\"{o}nrich \& Binney 2009).  However, we currently lack an understanding of the efficiency of migration, and this could be tested by large
surveys of elemental abundances.

\section{Concluding remarks}

I have only two `take-home' points:
\begin{itemize}

\item Spectroscopy is critical in determining the astrophysics of stellar populations

\item  Large (planned) imaging surveys need matched spectroscopic capabilities

\end{itemize}

\acknowledgements I thank the organisers for their invitation to this stimulating conference.  I acknowledge support from the US National Science Foundation  (AST-0908326).

\newpage%%%%%%%%%%%%%%%%%%%%%%%%%%%%%%%%%%%%%%%%%%%%%%%%%%%%%%

\end{document}